\documentclass[12pt]{article}
\usepackage{geometry}
\usepackage{a4}
\usepackage{graphicx}
\usepackage{epsf}
\usepackage{amsmath}
\usepackage{amssymb}
\usepackage{cite}
\newcommand{\be}{\begin{equation}}
\newcommand{\ee}{\end{equation}}

\newcommand{\Rmnum}[1]{\expandafter\@slowromancap\romannumeral #1@}
\newcommand{\bea}{\begin{eqnarray}}
\newcommand{\eea}{\end{eqnarray}}

\begin{document}
\def\C{{\mathbb{C}}}
\def\R{{\mathbb{R}}}
\def\s{{\mathbb{S}}}
\def\T{{\mathbb{T}}}
\def\Z{{\mathbb{Z}}}
\def\W{{\mathbb{W}}}
\def\Bbb{\mathbb}
\def\BZ{\Bbb Z} \def\BR{\Bbb R}
\def\BW{\Bbb W}
\def\BM{\Bbb M}
\def\BC{\Bbb C} \def\BP{\Bbb P}
\def\CP{\BC\BP}
\begin{titlepage}
\title{On Black Hole Entropy Corrections in the Grand Canonical Ensemble}
\author{}
\date{
Subhash Mahapatra, Prabwal Phukon, Tapobrata Sarkar
\thanks{\noindent E-mail:~ subhmaha, prabwal, tapo @iitk.ac.in}
\vskip0.4cm
{\sl Department of Physics, \\
Indian Institute of Technology,\\
Kanpur 208016, \\
India}}
\maketitle
\abstract{
\noindent
We study entropy corrections due to thermal fluctuations for asymptotically AdS black holes in the grand canonical ensemble. To leading 
order, these can be expressed in terms of the black hole response coefficients via fluctuation moments. We also analyze 
entropy corrections due to mass and charge fluctuations of R-charged black holes, and our results indicate an universality in the logarithmic corrections to 
charged AdS black hole entropy in various dimensions. }
\end{titlepage}

\section{Introduction}

Black holes are probably the most mysterious objects in the General Theory of Relativity, that are still far from being fully understood. Whereas a quantum
description of the nature of black hole spacetimes has remained elusive as of now apart from certain specific examples in loop quantum gravity (LQG), a lot of progress has 
nevertheless been made in the understanding of black hole thermodynamics (see, e.g \cite{wald},\cite{page}). Although the origin of entropy in black holes is quantum in nature 
unlike ordinary thermodynamic systems, semi classical analyses have often provided insights into the rich phase structure possessed by black hole systems. 
For example, it is well known by now that, whereas asymptotically flat black holes are 
thermally unstable, asymptotically AdS ones can be in thermal equilibrium with their own radiation. These often exhibit critical phenomena akin to phase transitions in ordinary 
liquid gas systems. Phase structures for AdS black holes have been extensively studied in the literature \cite{chamblin1},\cite{chamblin2},\cite{tapo1} and for many cases, critical 
exponents have been computed \cite{lousto},\cite{cai1},\cite{tapo2} revealing universality in a class of examples. 

It has been established that black hole entropy, as given by the celebrated Beckenstein-Hawking area law $S = \frac{A}{4}$ (in units of the Planck area) acquires
logarithmic corrections due to fluctuations in the black hole parameters, both at the quantum as well as the classical level. The former was studied in \cite{cm1} and the latter 
has been extensively investigated by several authors (see, e.g \cite{dmb},\cite{gour},\cite{cm2}).  In the canonical ensemble, where the system is in 
equilibrium with a heat bath and exchanges energy with the surrounding, the entropy correction can be simply related to the specific heat of the black hole \cite{dmb}. This follows 
from standard analyses in thermodynamic fluctuation theory. In the grand canonical ensemble, where one allows for fluctuations in all the extensive thermodynamic 
parameters of the black hole, the situation is more intricate. Electrically charged black holes in the grand canonical ensemble have been studied for the four dimensional
RN-AdS black hole in \cite{cm2}, via the grand canonical partition sum, by postulating a power law relation between the energy, area and charge. It is interesting to consider 
generic examples where one allows for fluctuations in the angular momentum as well. 

The main idea in this paper is  to exploit the Smarr type relations between the black hole parameters to calculate entropy corrections, via fluctuation theory. This can also
take into account fluctuations of the angular momentum for rotating black holes, which, to the best of our knowledge have not been dealt with so far. 
Our method elucidates the general procedure to evaluate these corrections in the grand canonical 
ensemble in the presence of arbitrary fluctuating parameters in any dimension, using standard thermodynamics. Using this, we also calculate the entropy corrections in the mixed 
ensembles \cite{tapo1} of KN-AdS black holes in four dimensions, which allow for interesting phase coexistence behaviour. We further relate these corrections to black hole 
response coefficients. Our methods are applicable to string theoretic black holes as well, and we show that there exists a certain universality in the entropy corrections for 
electrically charged AdS black holes. 

This paper is organised as follows. In section 2, we review the general method to calculate fluctuation induced corrections to black hole entropy.  A formula
for such corrections, involving the black hole response coefficients is also written down, generalising the known canonical example result. This is illustrated
for the example of the BTZ black hole in section 3. In section 4, entropy corrections are evaluated in the grand canonical ensemble for AdS black holes in four dimension. In 
addition, we calculate these in the mixed ensembles alluded to in \cite{tapo1}. In section 5, we calculate entropy corrections in string theoretic black holes. 
Finally, we end with our conclusions in section 6.

\section{Black Hole Entropy and Thermal Fluctuations}

In order to calculate corrections to black hole entropy from thermal fluctuations, one starts from the expression for the partition function in the grand canonical ensemble, 
which is given, for continuous values of the energy and the charges by 
\begin{equation}
Z \left(\mu_i,\beta\right) =  \int_0^{\infty}\int_0^{\infty}\cdots \int_0^{\infty}\rho\left(E,N_i\right){\rm exp}\left[-\beta\left( E - \mu_iN_i\right)\right]dE dN_1dN_2\cdots dN_i
\label{one}
\end{equation}
where $\rho$ is the density of states, $N_i$ are the ``particle numbers," or ``charges,'' with $\mu_i$ being the corresponding chemical potentials. $\beta$ is the inverse temperature. 
In the canonical ensemble, the expression simplifies as $\rho$ is a function of only the energy of the system, and the integrations over the $N_i$s are absent. For 
charged rotating black holes,
$i=2$ with $N_1$ and $N_2$ being the electric charge and the angular momentum, and $\mu_1, \mu_2$ are the electric potential and angular velocity. 
Note that this expression is valid only for a continuous distribution of the eigenvalues of the 
Hamiltonian, the charge operator, the angular momentum operator etc. in the corresponding quantum theory. Eq.(\ref{one}) is then inverted \cite{bohr} by using an inverse 
Laplace transform to obtain the density of states, which, in the saddle point approximation, is
\begin{equation}
\rho\left(E,N_i\right) = \frac{Z\left(\beta_0, \lambda_{i0}\right){\rm exp}\left(\beta_0 E + \lambda_{i0}N_i\right)}{\left(2\pi\right)^{\frac{3}{2}} \Delta_0^{\frac{1}{2}}}
\label{two}
\end{equation}
where $\beta=\frac{1}{T}$ and $\lambda_i = -\frac{\mu_i}{T}$. By definition, $\lambda_i = \left(\frac{\partial S}{\partial N_i}\right)$, where $S$ is expressed as a function
of $\left(E,N_i\right)$. $\Delta_0$ is the determinant of the matrix 
\begin{equation}
D = 
\begin{pmatrix}
\frac{\partial^2{\rm ln}Z}{\partial \lambda_1^2} & \frac{\partial^2{\rm ln}Z}{\partial \lambda_1\partial\lambda_2} & -\frac{\partial^2{\rm ln}Z}{\partial \lambda_1\partial\beta}\\
\frac{\partial^2{\rm ln}Z}{\partial \lambda_2\partial\lambda_1} & \frac{\partial^2{\rm ln}Z}{\partial \lambda_2^2} &  -\frac{\partial^2{\rm ln}Z}{\partial \lambda_2\partial\beta}\\
-\frac{\partial^2{\rm ln}Z}{\partial \beta\partial\lambda_1} & -\frac{\partial^2{\rm ln}Z}{\partial \beta\partial\lambda_2} & \frac{\partial^2{\rm ln}Z}{\partial \beta^2}\\
\end{pmatrix}
\label{three}
\end{equation}
evaluated at the equilibrium values of the temperature and the chemical potentials. It is to be noted \cite{callen} that the logarithm of the (grand canonical) partition function appearing
in eq.(\ref{three}) is the generalised Massieu transform of the entropy, 
\begin{equation}
{\rm ln}Z = S - \beta E - \lambda_iN_i
\label{massieu}
\end{equation}
where the left hand side of eq.(\ref{massieu}) should be thought of as a function of $\beta$ and $\lambda$. 

It follows from eq.(\ref{massieu}) that the numerator of eq.(\ref{two}) is the exponential of
the grand canonical entropy (as a function of $E$ and $N_i$), and hence, we have, apart from irrelevant constants, 
\begin{equation}
S_g = {\rm ln}\rho + \frac{1}{2}{\rm ln}\Delta_0
\label{four}
\end{equation}

If we define the microcanonical entropy as the logarithm of the density of states (as a function of only energy), then the above formula gives a correction 
to the microcanonical entropy in the grand canonical (or canonical) ensemble. \footnote{Such corrections to the microcanonical entropy appear for ordinary 
thermodynamic systems as well, but for such systems, in the infinite volume limit, when one talks about thermodynamic quantities per unit volume, these can be neglected. }

The expression for $\Delta_0$ can be computed via the response coefficients of the black hole. For example, in the canonical ensemble where the system is in
equilibrium with a heat bath, eq.(\ref{four}) simplifies to 
\begin{equation}
S_c = {\rm ln}\rho + \frac{1}{2}{\rm ln}CT^2
\label{five}
\end{equation}
where $C$ is the specific heat and $T$ the equilibrium temperature. In the grand canonical ensemble, where we allow for the fluctuations in the particle numbers, 
the elements of the matrix $D$, which are
directly related to the moments of fluctuation of the black hole parameters (see, eg. \cite{callen}) get related to the black hole response coefficients. 
For example, in a grand canonical ensemble with the mass $M$ and electric charge $Q$ being the fluctuating parameters, it can be shown that the determinant $\Delta_0$ can be
written as
\begin{equation}
\Delta_0 = <\Delta M^2><\Delta Q^2> - \left(<\Delta M\Delta Q>\right)^2
\end{equation}
A similar formula holds for angular momentum fluctuations. Simple algebraic manipulations then show that
when the black hole is in a grand canonical ensemble with the energy and a single charge (electric charge or angular momentum) being allowed to fluctuate, the determinant
$\Delta_0$ can be expressed as \footnote{A similar relation can be readily obtained for more than two fluctuating parameters. The result however is lengthy, and not
particularly illuminating.} 
\begin{equation}
\Delta_0 = T^3 C_{\lambda}\kappa+T^4\lambda\alpha\kappa-T^4\alpha^2 
\label{master}
\end{equation}
where we have defined the heat capacity at constant $\lambda$, and the response coefficients $\kappa$ and $\alpha$ (analogous to the ``compressibility''
and the ``expansivity'') with  $N = \{Q,J\}$, as 
\begin{equation}
C_{\lambda}=\frac{1}{\beta}\left(\frac{\partial S}{\partial T}\right)_{\lambda},~~~\kappa=-\frac{1}{T}\left(\frac{\partial N}{\partial \lambda}\right)_T,
~~~\alpha=\left(\frac{\partial N}{\partial T}\right)_{\lambda T}
\label{response}
\end{equation}
This gives the general condition for stability in the grand canonical ensemble where one allows for fluctuations in the energy and a single charge, as 
\begin{equation}
C_{\lambda} > T\alpha\left(\frac{\alpha}{\kappa} - \lambda\right)
\label{stability}
\end{equation}
Eq.(\ref{stability}) generalises the condition for the positivity of the specific heat in the canonical ensemble.

On physical grounds, one would expect that eq.(\ref{stability}) holds in the regions where the black hole is globally stable (i.e the Gibbs free energy is negative). This is indeed the case 
for the BTZ black hole in the grand canonical ensemble, as this is locally as well as globally stable everywhere. \footnote{We allude to the standard definitions that a thermodynamic
system is locally stable if the Hessian of its entropy does not develop negative eigenvalues. Global stability implies that the Gibbs free energy is negative.} 
However, for other black holes, as we will elucidate in the sequel, the region of global stability is more constrained than 
the region of parameter space for which eq.(\ref{stability}) is valid. The latter equation can be shown to be valid when the black hole is locally stable. 

We also mention here that for the grand canonical ensemble, it is more appropriate to deal with (fixed) thermodynamic potentials than the charges. In this sense, eq.(\ref{master}) 
should be thought of as being expressed in terms of the appropriate potentials. The results that we will obtain will be in terms of the fixed potentials, which can then be
used to deduce leading order corrections to black hole entropy in the limit of small potentials. It is also to be kept in mind that our analysis is strictly valid only in the classical
regime, i.e for large black holes, away from extremality. 

Finally, note that equation (\ref{four}) is based on the assumption that the spectrum of energy and the charges are continuous parameters. The situation changes markedly if this is not
the case (as is expected from LQG). We should, therefore, include an appropriate Jacobian in going to the continuum limit in eq.(\ref{one}) \cite{gour}. 
The Jacobian factor \footnote{We denote the Jacobian by $K$ in this paper, so as not to confuse with the notation for angular momentum} $K$
modifies the result for the entropy correction, and the corrected entropy is given by 
\begin{equation}
S_g = {\rm ln}\rho + {\rm ln}K + \frac{1}{2}{\rm ln}\Delta_0
\label{jacobian}
\end{equation}
Evaluation of the Jacobian factor requires knowledge about the quantum spectrum of the black hole parameters. This is not fully understood, and we will
proceed with the assumption that the area, charge and angular momentum spectra are linear in the quantum numbers that they are measured in \cite{cm2}. 

Let us also mention that an equivalent approach is to calculate the partition function of eq.(\ref{one}) directly, in the Gaussian approximation \cite{cm1}. It can be checked that this 
gives the same result for the correction to the entropy as that outlined above. 

We now illustrate this procedure for the case of the BTZ black hole in the grand canonical ensemble. 

\section{BTZ Black Hole in the Grand Canonical Ensemble}

The entropy of the BTZ black hole is given, in terms of its mass and angular momentum as
\begin{equation}
S = \frac{4\pi}{\sqrt{2}} \left[Ml^2\left(1 + \left[1 - \frac{J^2}{M^2l^2}\right]^{\frac{1}{2}}\right)\right]^{\frac{1}{2}}
\label{six}
\end{equation}
where $l$ is related to the cosmological constant $\Lambda = -\frac{1}{l^2}$. A useful quantity for the thermodynamic description is the angular velocity, which is given by the expression
\begin{equation}
\Omega = 8\pi^2\frac{J}{S^2}
\label{seven}
\end{equation}
in terms of which the Hawking temperature of the black hole is 
\begin{equation}
T = \frac{1}{8\pi^2l^2}S\left(1 - \Omega^2l^2\right)
\label{eight}
\end{equation}
which implies that non extremal BTZ black holes with non zero temperature exist only for $\Omega l < 1$ \cite{hawking}. The correction to the entropy can be 
evaluated from eq.(\ref{three}) by noting that 
\begin{equation}
\beta = \frac{1}{T};~~~~~\lambda = \frac{8\pi^2\Omega l^2}{S\left(\Omega^2l^2 - 1\right)}
\label{nine}
\end{equation}
The determinant $\Delta_0$ can be calculated from eq.(\ref{three}) or eq.(\ref{master}) : 
\begin{equation}
\Delta_0 = \frac{1}{4096}\frac{S^6\left(1-\Omega^2 l^2\right)^3}{\pi^8l^6}
\end{equation}
For a continuous energy spectrum, the final result is, apart from unimportant constants,
\begin{equation}
S_g = {\rm ln}\rho + 3{\rm ln}S_{{\rm bh}} 
\label{ten}
\end{equation}
where $S_{{\rm bh}}$ is identified with the area of the BTZ black hole. 
Hence we see that in the grand canonical ensemble, the entropy correction is proportional to ${\rm ln}S$, but with a prefactor of $3$, instead of $\frac{3}{2}$ in the 
canonical ensemble, calculated in \cite{dmb},\cite{cm1}. Inclusion of the Jacobian factor in eq.(\ref{jacobian}) alters the result. 
Following \cite{cm2}, for the BTZ black hole, this is given by
\begin{equation}
K^{-1} ~=~ \left(\frac{\partial E}{\partial x}\right)\left(\frac{\partial J}{\partial y}\right) ~=~\left(\frac{\partial E}{\partial A}\right)\left(\frac{\partial A}{\partial x}\right)
\left(\frac{\partial J}{\partial y}\right)
\label{jac1}
\end{equation}
Assuming that the spectra of the area and the angular momentum are linear in the quantum numbers $x$ and $y$, we obtain, apart from multiplicative constants,
\footnote{For determining leading order corrections, it is enough for us to use the equilibrium values of the energy and the area.}
\begin{equation}
K = \frac{8l^2\pi^2}{S\left(1- \Omega^2l^2\right)}
\end{equation}
Inclusion of this factor thus implies that the final form of the corrected entropy for the BTZ black hole is 
\begin{equation}
S_g = {\rm ln}\rho + 2{\rm ln}S_{{\rm bh}}
\end{equation}
It is also to be noted that for the BTZ black hole,
\begin{equation}
\Delta_0 =  -\frac{G^3}{\pi^2}
\end{equation}
where $G$ is the Gibbs free energy given by 
\begin{equation}
G~=~M - TS - \Omega J
\end{equation}
This ensures that the determinant $\Delta_0$ is always positive in the region of global stability where $G$ is negative. 

Finally, a few words about the fluctuation moments. It follows from eq.(\ref{three}) that the relative fluctuations are
\begin{eqnarray}
\frac{<\Delta M^2>}{M^2} &=& \frac{4\left(1 + 3\Omega^2l^2\right)}{S\left(1 + \Omega^2l^2\right)}\nonumber\\
\frac{<\Delta J^2>}{J^2} &=& \frac{1 + 3\Omega^2l^2}{S\Omega^2l^2}\nonumber\\
\frac{<\Delta M\Delta J>}{MJ} &=& \frac{2\left(3 + \Omega^2l^2\right)}{S\left(1 + \Omega^2l^2\right)}
\label{flucsbtz}
\end{eqnarray}
Eq.(\ref{flucsbtz}) shows the qualitative difference between mass and angular momentum fluctuations for BTZ black holes. In the limit of small $\Omega l$, the relative
fluctuation of the mass $\frac{\Delta M^2}{M^2} \sim S^{-1}$, as does the cross correlation term. The relative angular momentum fluctuation however
goes as  $\frac{\Delta J^2}{J^2} \sim S^{-1}\left(\Omega l\right)^{-2}$

\section{$D = 4$ AdS Black Holes in Various Ensembles}

We now turn to the example of AdS black holes in four dimensions. The generalised Smarr formula for KN-AdS black holes is well known \cite{caldarelli}
\begin{equation}
m= \frac{\sqrt{s^2\pi^2+4\pi^4j^2+\pi^4q^4+2q^2\pi^3s+4j^2\pi^3s+2q^2s^2\pi^2+2s^3\pi+s^4} }{2\pi^{3/2}\sqrt{s} }
\label{eleven}
\end{equation}
Here, we have conveniently rescaled the entropy, mass, charge and angular momentum of the black hole by the AdS radius as
\begin{equation}
s = \frac{S}{l^2},~~~m = \frac{M}{l},~~~q = \frac{Q}{l},~~~j = \frac{J}{l^2}
\end{equation}
Equivalently, one could set the AdS radius $l$ to be unity and work with the unscaled parameters. \footnote{In this section and the next, we will use lower case letters to 
denote the thermodynamic parameters, with the understanding that these are scaled parameters in the sense just mentioned. This is done to simplify the algebra, and factors 
of the AdS radius can be included at any stage.} 

Let us begin with the Kerr-AdS black hole, obtained by setting $q=0$ in eq.(\ref{eleven}). In this
case, it is convenient to express the angular momentum in terms of the angular velocity \footnote{Note that the angular velocity that enters in the formulae is the 
one measured w.r.t a non rotating frame at infinity \cite{caldarelli},\cite{gibbons}, i.e it is the difference of the angular velocity at the horizon and that at the boundary 
of spacetime. } as 
\begin{equation}
j= \frac{\omega s^{3/2}\sqrt{s+\pi} }{2 \pi^{3/2}\sqrt{\pi+s-\omega ^2s}  }
\label{twelve}
\end{equation}
In terms of the angular velocity, the temperature can be expressed as 
\begin{equation}
t= \frac{\pi^2+4s\pi-2\pi\omega ^2s+3s^2-3s^2\omega ^2}{4\pi^{3/2}\sqrt{s(\pi+s)(\pi+s-\omega ^2s)} }
\label{thirteen}
\end{equation}
The Massieu transform of the entropy is obtained as 
\begin{equation}
{\rm ln} Z= \frac{s(s^2-s^2\omega ^2-\pi^2)}{\pi^2+4s\pi-2\pi\omega ^2s+3s^2-3s^2\omega ^2}
\label{fourteen}
\end{equation}
where, as usual, the right hand side of eq.(\ref{fourteen}) has to be thought of as a function of $\beta$ and $\lambda = -\beta\omega$, where
\begin{equation}
\beta = \frac{1}{t},~~~\lambda = -\frac{4\pi^{3/2}\omega\sqrt{s(s+\pi)[s(1-\omega ^2)+\pi]} } {3s^2(1-\omega ^2)+2s\pi(2-\omega ^2)+\pi^2}
\end{equation}
and $\omega$ is the angular velocity given by 
\begin{equation}
\omega = \frac{2\pi^{\frac{3}{2}}j\sqrt{s+\pi}}{\sqrt{s}\sqrt{s^3 + s^2\pi + 4j^2\pi^3}}
\end{equation}
The exact expression for the determinant in eq.(\ref{four}) (or equivalently eq.(\ref{master})) is somewhat lengthy and will not be reproduced here. We simply state that in the 
limit of small angular velocities, we obtain the result
\begin{equation}
\Delta_0 = \frac{s(\pi+3s)^4}{64\pi^6(3s-\pi)}
\label{fifteen}
\end{equation}
Hence, the corrected grand canonical entropy is, in this case, for continuous distributions of the area and angular momentum,
\begin{equation}
s_g = {\rm ln}\rho + 2{\rm ln}s_{{\rm bh}}
\label{sixteen}
\end{equation}
\begin{figure}[t!hbp]
\centering
\includegraphics[width=3in,height=2.5in]{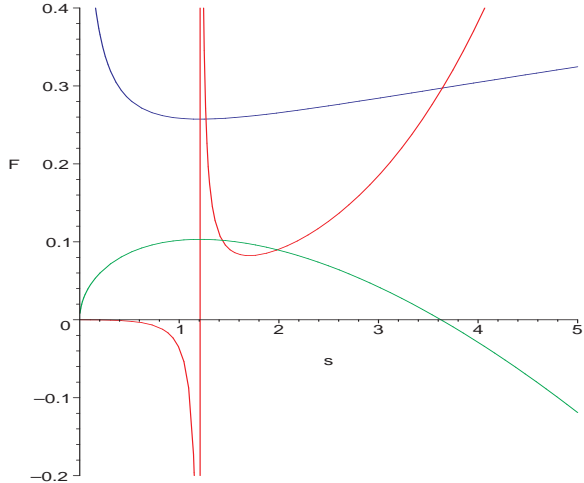}
\caption{Plots of various parameters (denoted by F) of the Kerr AdS black hole with entropy, for $\omega = \frac{1}{2}$. The blue curve denotes the temperature, the 
green denotes the Gibbs free energy and the red curve is for $\Delta_0$}
\label{kads}
\end{figure}

The relative fluctuations can be calculated for the KAdS black hole as in the BTZ example, and we find that similar to the latter, $\frac{<\Delta m^2>}{m^2}$ and 
$\frac{<\Delta m \Delta j>}{mj}$  goes as $s^{-1}$ whereas $\frac{<\Delta j^2>}{j^2}\sim s^{-1}\omega^{-2}$

Before moving on, a word about the stability of the system for different values of the black hole parameters. Calculation of the Gibbs free energy in this case shows that for a given 
value of $\omega$, the free energy becomes negative for $s = \pi\left(1 - \omega^2\right)^{-\frac{1}{2}}$. For $\omega = \frac{1}{2}$, this implies that $s = 3.63$. This signals the
Hawking-Page phase transition, with $t_{hp} = 0.297$. 
On the other hand, the Davies temperature $t_d$ of the black hole (where the specific heat diverges) can be calculated, for $\omega = \frac{1}{2}$ to be at 
$t_d = 0.257$ (for $s = 1.208$). This is the temperature for which $\Delta_0$ diverges as well, via a change of sign. Thus, positive regions of $\Delta_0$ denotes a locally
stable black hole, which remains metastable till the Hawking-Page point before becoming globally stable.  \footnote{A similar conclusion can be reached by calculating the
eigenvalues of the Hessian of the entropy, as in standard thermodynamics.} These results are graphically depicted in fig.(\ref{kads}), where the blue, green and red curves are the 
plots of the temperature, the Gibbs free energy and $\Delta_0$ for various values of the entropy. 

We now calculate the Jacobian factor of eq.(\ref{jacobian}). Assuming again that the spectra of the area and the angular momentum are linear in their respective quantum 
numbers, we obtain  
\begin{equation}
K = \frac{4\pi^{3\over 2}\sqrt{s\left(\pi + s\right)\left(\pi + s(1 - \omega^2)\right)}}{3s^2\left(1 - \omega^2\right) + 2\pi s\left(2 - \omega^2\right) + \pi^2}
\end{equation}
For small values of $\frac{\omega}{s}$, $K \sim s^{-\frac{1}{2}}$, and the final form of the corrected entropy in the grand canonical ensemble is
\begin{equation}
s_g = {\rm ln}\rho + \frac{3}{2}{\rm ln}s_{{\rm bh}}
\label{finalkads}
\end{equation}

A similar analysis can be carried out for RN-AdS black holes, obtained by setting $j=0$ in eq.(\ref{eleven}). We will simply state the main result here. Expressing the charge 
in terms of the potential as $q = \frac{\phi \sqrt{s} }{\sqrt{\pi}}$, we obtain
\begin{equation}
\lambda = -\frac{4\phi \pi^{3/2}\sqrt{s} }{3s+\pi(1- \phi^2)}
\end{equation}
the Massieu transform of the entropy, given by the logarithm of the grand canonical partition function, is 
\begin{equation}
{\rm ln}Z= \frac{s[s-\pi(1-\phi^2)]}{3s+\pi(1- \phi^2)}
\label{rnads1}
\end{equation}
where, as usual, the r.h.s of eq.(\ref{rnads1}) has to be considered as a function of $\beta$ and $\lambda$. Finally, $\Delta_0$ is given by
\begin{equation}
\Delta_0 = \frac{1}{32}\frac{\left(3s + \pi\left(1 - \phi^2\right)\right)^4}{\pi^5\left(3s - \pi\left(1 - \phi^2\right)\right)}
\label{seventeen}
\end{equation}
In the limit $s \gg \phi$, the correction to the entropy is 
\begin{equation}
s_g = {\rm ln}\rho + \frac{3}{2}{\rm ln}s_{{\rm bh}}
\label{eighteen}
\end{equation}
Whereas this is true for the case of continuous energy and charge spectra, inclusion of the Jacobian factor of eq.(\ref{jacobian}) modifies the result to
\begin{equation}
s_g = {\rm ln}\rho + {\rm ln}s_{{\rm bh}}
\label{finalrnads}
\end{equation}
as can be straightforwardly deduced from an analogue of eq.(\ref{jac1}). This is in agreement with the result obtained in \cite{gour}.

For the RN-AdS black hole in the grand canonical ensemble, one can calculate the relative fluctuations of the mass and charge as before. Here, we find that whereas
the relative fluctuation in the mass and the cross correlation term between the mass and the charge goes as $s^{-1}$, as for rotating black holes, the relative
fluctuation $\frac{<\Delta q^2>}{q^2} \sim \phi^{-2}$. This is different from the result obtained in \cite{cm2}, and needs to be investigated further. \footnote{Using $\alpha = 
\frac{3}{2}$ and $\beta = \frac{1}{2}$ in eq.(27) of \cite{cm2}, we seem to reach the same conclusion, i.e the relative fluctuation of the charge $\sim \phi^{-2}$}

Entropy corrections to the general KN-AdS black holes in four dimensions can be similarly determined, although the formulae are too long to reproduce here. We 
state the final result that in the limit of small potentials (electric potential or angular velocity), the entropy correction to the KN-AdS black hole is given by 
\begin{equation}
s_g = {\rm ln}\rho + \frac{5}{2}{\rm ln}s_{{\rm bh}}
\label{nineteen}
\end{equation}
for the case of continuous energy electric charge and angular momentum distributions. Including the Jacobian factor modifies the coefficient 
${5 \over 2}$ to $2$. The relative fluctuations of the angular momentum and the charge follow the same behaviour as in the KAdS and RNAdS cases. All the
cross correlation terms fall off as $s^{-1}$. The behaviour of the relative charge fluctuation, which again goes as $\frac{1}{\phi^2}$ 
needs to be studied further. \footnote{We postpone further discussions on this to the final section.}  

It is interesting to further study the KN-AdS black hole in the mixed ensembles \cite{tapo2} alluded to in the introduction. These are defined to be ensembles in which one
thermodynamic charge (i.e the electric charge or the angular momentum) and the potential conjugate to the other (the electric potential or the
angular velocity) are held fixed. We start with the fixed (electric) charge ensemble. In this
case, the angular momentum can be solved in terms of the angular velocity as
\begin{equation}
j= \frac{\omega(\pi s+s^2+\pi^2q^2)\sqrt{s}}{2\pi^{3/2}\sqrt{(\pi+s)(\pi+s-\omega^2s)} }
\label{twenty}
\end{equation}
The Massieu transform of the entropy yields the logarithm of the partition function in the fixed charge ensemble
\begin{equation}
{\rm ln} Z =\frac{s(s^4-s^4\omega^2+\pi s^3-\pi\omega^2 s^3+3\pi^2\omega^2s^2-4\pi^2s^2-7\pi^3s+2\pi^3\omega^2s-3\pi^4)} {3s^4-3s^4\omega^2+7s^3\pi-5\pi\omega^2s^3+4s^2\pi^2-\pi^2\omega^2s^2-\pi^3s-\pi^4}
\label{twentyone}
\end{equation}
The determinant $\Delta_0$ (for the case $q=1$) is given, in the limit of small $\omega$ by
\begin{equation}
\Delta_{0,q=1}= \frac{(s^2+s\pi+\pi^2)(3s^2- s\pi-\pi^2)^4}{64\pi^6s^3(s+\pi)(3s^2-s\pi+3\pi^2)}
\end{equation}
from which it can be seen that the correction to the entropy follows the same equation as the Kerr-AdS black hole of eq.(\ref{finalkads}), as expected. 
The fixed angular momentum case can be similarly studied and the entropy correction is found to be the same as in eq.(\ref{finalrnads}).  

We end this section with a couple of comments about the mixed ensembles. It is known that in the fixed charge ensemble, the black hole exhibits 
phase coexistence of large and small black hole branches, and also shows first order phase transitions akin to Van der Waals systems (which is also seen in the canonical ensemble 
of the RN-AdS or Kerr-AdS black holes) \cite{tapo2}. This happens below a critical value of the (fixed) charge, i.e $q=\frac{1}{6}$. \footnote{The reader is referred to fig. (31) of
\cite{tapo2} for a quick reference.} Although we are mainly interested in the large black hole regime, it is nevertheless interesting to explore the behaviour of $\Delta_0$ in the regions 
of phase coexistence. We find that $\Delta_0$ remains positive in the physical regions of an isotherm (where the specific heat $c_{\omega}$ is positive), remains negative in the 
unphysical region (where the specific heat is negative), changing sign through divergences at the turning points of the isotherm (where the specific heat diverges). 
For $q > \frac{1}{6}$, no phase coexistence exist (for $\omega < 1$) and $\Delta_0$ is always positive, starting from zero at extremality. 
A similar analysis can be done for the fixed $j$ ensemble as well.

\section{R-charged Black holes in Various Dimensions}

Finally, we address the issue of entropy corrections in string theoretic black holes in dimensions $D=5, 7$ and $4$, corresponding to rotating $D3$, $M5$ and $M2$ branes
in the grand canonical ensemble. These are familiar examples of R-charged black holes \cite{bcs},\cite{cg}. For example, the $D=5$ case corresponds to a spinning 
$D3$-brane configuration in which rotations in planes orthogonal to the brane is characterized by the group $SO(6)$. Upon a Kaluza Klein reduction of the spinning 
$D3$-brane on $S^5$, the three independent spins on the $D3$-brane world volume (which are the three independent Cartan generators of $SO(6)$) reduce to three 
$U(1)$ gauge charges of the corresponding $AdS_5$ black hole. We will only deal
with the compact horizon case here, and our notations will follow \cite{cg} (see also \cite{tapo3}). The analysis proceeds 
entirely in the same manner as that outlined previously. For the single R-charged black hole in $D=5$ (case 1 of \cite{tapo3}), the mass is given by
\begin{equation}
m= \frac{3}{2}r_{+}^{4}+\frac{3}{2}r_{+}^{2}+\frac{3}{2}r_{+}^{2}a+a
\label{twentytwo}
\end{equation}
and the entropy is
\begin{equation}
s= 2\pi r_{+}^{2}\sqrt{r_{+}^{2}+a}
\end{equation}
where $r_+$ denotes the position of the horizon, the charge parameter $a$ is related to the physical charge by
\begin{equation}
q= \sqrt{a(r_{+}^{2}+a)(r_{+}^{2}+1)}
\end{equation}
It is useful to solve for the charge parameter $a$ in terms of the electric potential \cite{tapo3},
\begin{equation}
a= \frac{r_{+}^{2}\phi^2}{r_{+}^{2}+(1-\phi^2)}
\end{equation}
Then, the calculation of the Massieu transform of the entropy is standard, and this is given by
\begin{equation}
{\rm ln}Z = \frac{\pi r_{+}^{3}(r_{+}^{4}-1+\phi^2)}{(2r_{+}^{2}+1-\phi^2)\sqrt{r_{+}^{2}+1}\sqrt{r_{+}^{2}+1-\phi^2}}
\end{equation}
from this, we can calculate, as before, 
\begin{equation}
\Delta_0 =  \frac{r_{+}^{2}(r_{+}^{2}+1)^3(2r_{+}^2+1-\phi^2)^4(3r_{+}^4+6r_{+}^2+3+\phi^2) }{4\pi^2(r_{+}^2+1-\phi^2)^3[2r_{+}^6+3r_{+}^4(1-\phi^2)-1+2\phi^2-\phi^4]}
\end{equation}
In the grand canonical ensemble, where $\phi$ is fixed to a small value compared to the horizon radius, $s \sim r_+^3$, and $\Delta_0 \sim r_+^8$, and hence
\begin{equation}
s_g = {\rm ln}\rho + {\rm ln}s_{{\rm bh}}
\label{twentythree}
\end{equation}
where we have used the fact that the Jacobian of eq.(\ref{master}) goes, in this case, as $K \sim s^{-\frac{1}{3}}$.
For single R-charged black holes in $D=4$ and $7$, we find that the correction to the entropy follows the same rule as in eq.(\ref{twentythree}), indicating an
universality in the entropy correction for such black holes in various dimensions. Note further that this is the same as eq.(\ref{finalrnads}). This result indicates that
charged AdS black holes in any dimensions have an universal logarithmic correction to the entropy, with unit coefficient. Such universality is however apriori not
obvious for rotating AdS black holes.

\section{Discussions and Conclusions}

In this paper, we have studied entropy corrections due to thermal fluctuations for a wide class of AdS black holes in the grand canonical ensemble and a couple of
mixed ensembles. We have argued that the Smarr formula
can be effectively used to calculate such corrections, by using standard tools of thermodynamics, and are expressible entirely in terms of the black hole response coefficients. 
The generalised Massieu transform of the entropy can also
be used to calculate moments of charge, mass and angular momentum fluctuations. Our results highlight the difference between charge and angular momentum 
fluctuations. Such differences are also seen in the study of the state space scalar curvature of thermodynamic geometries \cite{tapo2}. As mentioned in the text,
the case of relative charge fluctuations in the grand canonical ensemble needs to be investigated further. For the RN-AdS as well as the KN-AdS black hole, we find that
this is proportional to the inverse of the potential. For R-charged black holes, $\frac{<\Delta m^2>}{m^2}$ and $\frac{<\Delta m\Delta q>}{mq} \sim s^{-1}$ 
in $D = 5, 7$ and $4$. Interestingly, whereas for $D=5$ and $7$, the relative fluctuation  $\frac{<\Delta q^2>}{q^2}$ falls off as a fractional power of $s$, in $D=4$, we find
$\frac{<\Delta q^2>}{q^2} \sim \frac{1}{\phi^2}$. We leave a discussion on this for a future work. 

In this work, we have ignored the quantum corrections to the black hole entropy \cite{kaul} as appears in the microcanonical ensemble. It should be possible, however,
to include such corrections in our analysis in known examples. It would also be interesting to understand the role of thermal fluctuations for multiply charged
string theoretic black holes in the canonical ensemble, especially since the latter exhibit rich phase structure and liquid gas like first order phase transitions. 
Finally, the role of thermal fluctuations in string theoretic black holes should lead to interesting results in the dual gauge theory side initially studied in \cite{sudipto}. This
is left for a future investigation. 

\begin{center}
{\bf Acknowledgements}
\end{center}
It is a pleasure to thank S. Das, P. Majumdar and V. Subrahmanyam for several useful discussions and clarifications. The work of SM is supported by grant 
no. 09/092(0792)-2011-EMR-1 of CSIR, India.

\end{document}